\documentclass[lettersize,journal]{IEEEtran}
\usepackage{amsmath,amsfonts}
\usepackage{algorithm}
\usepackage{array}
\usepackage[caption=false,font=normalsize,labelfont=sf,textfont=sf]{subfig}
\usepackage{textcomp}
\usepackage{stfloats}
\usepackage{url}
\usepackage{verbatim}
\usepackage{graphicx}
\def\BibTeX{{\rm B\kern-.05em{\sc i\kern-.025em b}\kern-.08em
    T\kern-.1667em\lower.7ex\hbox{E}\kern-.125emX}}
\usepackage{balance}

\usepackage{amssymb}
\usepackage{booktabs} 
\usepackage{algpseudocode}
\usepackage{makecell}
\usepackage{enumerate}
\usepackage{cite}

\begin{document}

\title{Iterative Equalization of CPM With Unitary Approximate Message Passing}

\author{Zilong Liu, Yi Song, Qinghua Guo, \textit{Senior Member, IEEE}, Peng Sun, Kexian Gong, Zhongyong Wang
\thanks{Z. Liu, P. Sun, K. Gong and Z. Wang are with the School of Electrical and Information Engineering, Zhengzhou University, 450001, Zhengzhou, China (zilongliu@gs.zzu.edu.cn; iepengsun@zzu.edu.cn; kxgong@zzu.edu.cn; zywangzzu@gmail.com;).} 
\thanks{Y. Song is with the Key Laboratory of Grain Information Processing and Control, Ministry of Education, the College of Information Science and Engineering, Henan University of Technology, Zhengzhou 450001, China (syczzu@gmail.com);} 
\thanks{Q. Guo is with the School of Electrical, Computer and Telecommunications Engineering, University of Wollongong, Wollongong, NSW 2522, Australia (qguo@uow.edu.au).}
}


\maketitle

\begin{abstract}
Continuous phase modulation (CPM) has extensive applications in wireless communications due to its high spectral and power efficiency. However, its nonlinear characteristics pose significant challenges for detection in frequency selective fading channels. This paper proposes an iterative receiver tailored for the detection of CPM signals over frequency selective fading channels. This design leverages the factor graph framework to integrate equalization, demodulation, and decoding functions. The equalizer employs the unitary approximate message passing (UAMP) algorithm, while the unitary transformation is implemented using the fast Fourier transform (FFT) with the aid of a cyclic prefix (CP), thereby achieving low computational complexity while with high performance. For CPM demodulation and channel decoding, with belief propagation (BP), we design a message passing-based maximum a posteriori (MAP) algorithm, and the message exchange between the demodulator, decoder and equalizer is elaborated. With proper message passing schedules, the receiver can achieve fast convergence. Simulation results show that compared with existing turbo receivers, the proposed receiver delivers significant performance enhancement with low computational complexity. 
\end{abstract}


\section{Introduction}
\IEEEPARstart{C}{ontinuous} phase modulation (CPM) plays an important role in the field of wireless communications. The primary advantages of CPM signals include their constant envelope characteristics and continuous phase transitions \cite{Ref1}. These features provide robust resilience against nonlinear amplification effects, thereby enabling increased flexibility in power amplifier design. Furthermore, the spectral compactness of CPM signals signifies enhanced efficiency in bandwidth utilization \cite{Ref2, Ref3}, an attribute of particular importance in environments where spectrum resources are scarce. Consequently, CPM has been extensively used in mobile communications, satellite communications, aerospace, the internet of things, and integrated sensing and communication fields\cite{Ref4,Ref5,Ref6,Ref7}. \par 

However, the superior characteristics of CPM signals also bring challenges in receiver design. The multiplicity of modulation parameters and the classification of CPM signals as memory-dependent modulations necessitate more intricate processing techniques \cite{Ref8,Ref9}. Typically, maximum likelihood sequence detection (MLSD) or maximum a posteriori processing (MAP) algorithms are employed, requiring the use of an extensive array of matched filters at the receiver end and the use of Viterbi \cite{Ref10} or BCJR \cite{Ref11} algorithms for subsequent processing. In multipath fading channels, CPM signals are substantially impacted by intersymbol interference (ISI), which further complicates signal processing. The most effective receiver for CPM demodulation in such channels is believed to be the joint equalization and demodulation receiver \cite{Ref12}. However, the computational complexity of this algorithm increases exponentially with the channel length and the memory span of the signal. In order to reduce the complexity of the CPM signal processing, it is conventional practice to disentangle the equalizer from the detector \cite{Ref13,Ref14,Ref15,Ref16,Ref17}. This involves initiating with linear equalization on the received signal, followed by demodulation of the CPM signal using Viterbi or BCJR algorithms. Although suboptimal, this method significantly alleviates computational complexity.\par

Turbo equalization is adept at mitigating ISI, achieving enhanced performance through iterative processing between the equalizer and decoder \cite{Ref18,Ref19,Ref20}. The initial implementations of turbo equalization comprised two trellis-based detectors \cite{Ref18}, with one specifically assigned to equalization and the other to channel decoding. This facilitated the exchange of extrinsic information via the BCJR algorithm. Subsequently, \cite{Ref19} integrated linear filter-based equalizers with signal decoding techniques, attaining performance similar to trellis-based algorithms while substantially reducing computational complexity. However, the aforementioned turbo equalization methods are applicable to linear modulation signals and cannot be directly applied to nonlinear CPM signals. Many works have been dedicated to the turbo equalization of CPM signals specifically \cite{Ref21,Ref22,Ref23,Ref24,Ref25,Ref26}. \par 

In \cite{Ref21}, a turbo linear equalizer (TLE) based on Laurent decomposition for CPM was introduced. However, the algorithm has high complexity and the turbo gain is not significant. In contrast, \cite{Ref22} proposed a dual-iterative equalizer that integrates equalization, CPM demodulation, and channel decoding based on the minimum mean square error (MMSE) with soft interference cancellation (SIC) algorithm. \cite{Ref22} changes the number of back-end iterations after each front-end iteration to improve the quality of the prior information for the next front-end iteration, thereby improving performance and accelerating convergence. In \cite{Ref23}, the application of the dual-iterative algorithm was extended to the frequency domain, achieving similar performance gains while reducing complexity. Similarly, in \cite{Ref24}, an MMSE/SIC-based filtering technique was proposed to eliminate the self-interference caused by different components of the Laurent decomposition. In \cite{Ref25}, the study used an unknown channel and least squares (LS) for channel estimation, followed by an MMSE equalizer for turbo equalization. The above works \cite{Ref21,Ref22,Ref23,Ref24,Ref25} are all based on the MMSE, which reduces the equalizer performance because each transmitted symbol is considered as a Gaussian variable. The non-coherent detection algorithm described in \cite{Ref26} avoids the shortcomings associated with the MMSE algorithm. \cite{Ref26} utilized the per-survivor processing (PSP) algorithm to reduce the memory span of both the channel and the CPM, which can reduce the complexity of the trellis-based equalizer algorithm. However, the complexity remains a significant challenge compared to MMSE-based algorithms. \par

Message passing based turbo equalization has been used in systems with linear modulations, resulting in highly commendable outcomes \cite{Ref27,Ref28,Ref29}. Furthermore, in complex systems, adopting joint message passing rules can often bring even more significant advantages \cite{Ref30}. However, there is a scarcity of literature applying the message passing to the processing of CPM signals, with the majority of studies limited to the belief propagation (BP) algorithm \cite{Ref31,Ref32}. In this paper, we propose an iterative equalization algorithm for CPM based on message passing on a factor graph. We first use the unitary approximate message passing (UAMP) algorithm \cite{Ref33} to handle the densely connected part of the factor graph, which corresponds to the equalizer. In particular, the unitary transformation is implemented using the fast Fourier transform (FFT) with the aid of a cyclic prefix (CP), thereby achieving low computational complexity while with high performance. Then, BP is used to deal with the demodulator and decoder, and a message passing schedule is designed to select different inner iterations to achieve a balance between complexity and performance. Finally, the effectiveness of the proposed algorithm is verified through simulations. The results show that, with low complexity, the proposed algorithm significantly outperforms the MMSE/SIC-based equalization algorithms in the literature. \par

The remainder of this paper is organized as follows. Section II introduces the model of CPM signals and the received signals. In Section III, we derive the UAMP-based iterative equalization algorithm based on the factor graph. The performance of the proposed algorithm is evaluated through simulations in Section IV. Finally, the paper is concluded in Section V.\par

\textit{Notations:} Bold lowercase letters denote vectors, while bold uppercase letters denote matrices. The superscript $\mathbf{A}^H$ denotes the conjugate transpose of a matrix, and $\mathbf{A}^T$ denotes the transpose. The notation $\mathbf{a}_n$ denotes the $n$-th subvector of $\mathbf{a}$, with $a_n^i$ specifically referring to the $i$-th element within $\mathbf{a}_n$. Similarly, $\mathbf{A}_n$ denotes the $n$-th submatrix of the matrix $\mathbf{A}$, and $\mathbf{A}_n^i$ specifically corresponds to the $i$-th row of the matrix $\mathbf{A}_n$. A complex Gaussian random variable $x$ with mean $\hat x$ and variance $v_x$ is expressed as $\mathcal{CN}(x;\hat{x},v_x)$. When a function $f(x)$ is proportional to $g(x)$ by a positive constant $c$, this relationship is denoted as $f(x)\propto g(x)$. The bold symbols $\mathbf{1} $ and $\mathbf{0}$ represent the vector of ones and zeros, respectively. Finally, $\text{Diag}(\mathbf{a})$ constructs a diagonal matrix whose diagonal elements are given by the vector $\mathbf{a}$.

\section{System model}
\begin{figure*}
    \centering
    \includegraphics[width=0.75\linewidth]{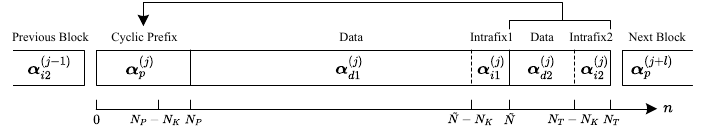}
    \caption{Data block structure of $\boldsymbol{\alpha}^{(j)}$ in the case of cyclic prefix.}
    \label{fig:1}
\end{figure*}

\subsection{CPM Signal}
A finite sequence of bits denoted as $\mathbf{b}=[b_0,b_1,...,b_{K-1}]^T$, is transmitted and processed through a convolutional code with a coding rate of $R$. The bit sequence is then interleaved to generate a codeword $\mathbf{c}=[\mathbf{c}_0,\mathbf{c}_1,...,\mathbf{c}_{N-1}]^T$.  Subsequently, the codeword $\mathbf{c}$ is mapped to form a sequence of CPM transmit symbols, denoted as $\boldsymbol{\alpha}=[\alpha_0,\alpha_1,...,\alpha_{N-1}]^T$.  Each symbol $\alpha_n$ corresponds to the $n$-th sub-vector $\mathbf{c}_n=[c_n^1,c_n^2,...,c_n^G]^T$ and takes on values from the set $\{\pm1,\pm3,...,\pm{M-1}\}$. Here, $M$ represents the modulation order, and $G=\log_2{M}$ represents the number of bits per transmitted symbol $\alpha_n$. The CPM baseband signal can be expressed as  
\begin{equation}
    \mathbf{x}(t,\boldsymbol{\alpha})=\sqrt{\frac{E_s}{T}}e^{j\varphi(t,\boldsymbol{\alpha})},
\end{equation}
where $E_s$ represents the symbol energy and $T$ denotes the symbol period. The phase component $\varphi(t,\boldsymbol{\alpha})$, which carries the modulation information is given as
\begin{equation}   \varphi(t,\boldsymbol{\alpha})=2\pi{h}\sum_{i=0}^{N-1}\alpha_ig(t-iT),
\label{equ:2}
\end{equation}
where $h=Q/P$ ($Q$ and $P$ being coprime integers) represents the modulation index. Additionally, the function $g(t)$ denotes the phase shaping, is detailed as 
\begin{equation}
    g(t)\triangleq\int_{-\infty}^t\omega(\tau)d\tau=\begin{cases}
            0 &t<0 \\
            \frac{1}{2}        & t\geqslant LT, 
         \end{cases}
\end{equation}
where $\omega(t)$ represents the frequency pulse function over a duration of $LT$,  which can be derived from a range of pulse shapes including the raised cosine (RC), rectangular (REC), or Gaussian minimum shift keying (GMSK) pulse. \par
The demodulation of CPM signals is dependent on the state trellis diagram. For time intervals within $t\in[nT,(n+1)T]$,  \eqref{equ:2} can be reformulated as follows
\begin{equation}
    \begin{split}
            \varphi(t,\boldsymbol{\alpha}) &=\pi{h}\sum_{i=0}^{n-L}\alpha_i+2\pi{h}\sum_{i=n-L+1}^n\alpha_ig(t-iT) \\
                &\triangleq\theta_n+\theta(t,\boldsymbol{\alpha}),
    \end{split}
    \label{equ:4}
\end{equation}
where $\theta_n=[\pi h\sum_{i=0}^{n-L}\alpha_i]\text{mod}2\pi$ represents the accumulation phase, which can assume $P$ distinct values for even $Q$, or $2P$ values for odd $Q$. The term $\theta(t,\boldsymbol{\alpha})=2\pi{h}\sum_{i=n-L+1}^n\alpha_ig(t-iT)$ represents the instantaneous phase, offering  $M^L$ potential variations. The phase trellis of CPM exhibits time-variant when $\theta_n$  assumes $2P$ values. To achieve a time-invariant phase trellis,  the tilted phase was introduced in \cite{Ref34}, formulated as 
\begin{equation}
    \phi(t,\boldsymbol{\alpha})=\varphi(t,\boldsymbol{\alpha})+\frac{\pi h(M-1)t}{T}.
\end{equation}
The corresponding tilted phase CPM signal is expressed as 
\begin{equation}   
    y(t,\boldsymbol{\alpha})=x(t,\boldsymbol{\alpha})e^{j\pi h(M-1)t/T}.
    \label{equ:6}
\end{equation}
With this transformation, the number of possible states for $\theta_n$ is consistently limited to $P$, irrespective of the values of $Q$.

The state at the $n$-th instance of the CPM signal is defined as
\begin{equation}
    \epsilon_n=\{\theta_n,\alpha_{n-1},...,\alpha_{n-L+1}\}.
\end{equation}
Hence, by applying the modulation relationship specified in \eqref{equ:4}, the phase of the signal during the interval $[nT, (n+1)T]$ can be derived from the current state $\epsilon_n$ and the incoming symbol $\alpha_n$. This corresponds to $PM^L$ potential outcome for $y(t,\boldsymbol{\alpha})$ at time $t\in[nT,(n+1)T]$, which are elements of the set 
\begin{equation}
    \boldsymbol{\chi}=\{\boldsymbol{\chi}_0,\boldsymbol{\chi}_1,...,\boldsymbol{\chi}_{PM^L-1}\}.
\end{equation}\par
Simultaneously, the subsequent state $\epsilon_{n+1}$ may also be determined
\begin{equation}
    \epsilon_{n+1}=\{\theta_{n+1},\alpha_{n},...,\alpha_{n-L+2}\},
\end{equation}
with $\theta_{n+1}$ being determinable through a recursive computational process 
\begin{equation}
    \theta_{n+1}=[\theta_n+2\pi h\tilde{\alpha}_{n-L+1}]\text{mod}2\pi,
\end{equation}
where 
\begin{equation}
    \tilde\alpha_i=\frac{\alpha_i+(M-1)}{2},
\end{equation}
is called modified data and takes on values within the set $\{0,1,...,M-1\}$. 

\subsection{Block-based CPM }
To facilitate equalization in the frequency domain, it is essential to insert a CP for each data block. Due to the inherent memory of CPM signals, in addition to ensuring that the length $N_P$ of the CP exceeds the memory length $L_c$ of the channel, two additional intrafixes of length $N_K$ must be inserted \cite{Ref35,Ref36}. As shown in Fig. \ref{fig:1}, the procedure within the $j$-th transmission data block consists of partitioning the data symbol $\boldsymbol{\alpha}$, spanning $N$ units, into two segments: $[\boldsymbol{\alpha}_{d1}^{(j)};\boldsymbol{\alpha}_{d2}^{(j)}]$, where the superscript $(j)$ represents the $j$-th block. Subsequently, two intrafixes $\boldsymbol{\alpha}_{i1}^{(j)}$ and $\boldsymbol{\alpha}_{i2}^{(j)}$ of length $N_K$ are inserted after $\boldsymbol{\alpha}_{d1}^{(j)}$ and $\boldsymbol{\alpha}_{d2}^{(j)}$ , respectively,  to form a block $[\boldsymbol{\alpha}_{d1}^{(j)}; \boldsymbol{\alpha}_{i1}^{(j)};\boldsymbol{\alpha}_{d2}^{(j)};\boldsymbol{\alpha}_{i2}^{(j)}]$ of length $\tilde N=N + 2N_K$. Finally, the sequence $[\boldsymbol{\alpha}_{d2}^{(j)};\boldsymbol{\alpha}_{i2}^{(j)}]$ is positioned ahead of $\boldsymbol{\alpha}_{d1}^{(j)}$ as CP, forming blocks of size $N_T = \tilde N + N_P$. This is represented as 
\begin{equation}
    \boldsymbol{\alpha}^{(j)}=[\boldsymbol{\alpha}_p^{(j)};\boldsymbol{\alpha}_{d1}^{(j)};\boldsymbol{\alpha}_{i1}^{(j)};\boldsymbol{\alpha}_{d2}^{(j)};\boldsymbol{\alpha}_{i1}^{(j)}].
\end{equation}
After this sequence of operations, the CPM state is observed to be consistent at time instants  $n=0$, $n=\tilde N-1$, and $n=N_T-1$, thereby assuring both intra-block and inter-block phase coherence throughout the transmission process.  The minimum length of $N_K$ satisfying the above condition depends on $P$ \cite{Ref36}
\begin{equation}
    N_K\geqslant P-1.
\end{equation}

\subsection{Received Signal Model}
Due to the impact of ISI and the presence of additive white Gaussian noise (AWGN), the received signal can be represented as follows
\begin{equation}
    r(t)=\sum_{m=0}^{L_c-1}h_mx(t-mT,\boldsymbol{\alpha})+v(t),
\end{equation}
where $h_m$ denote the gain associated with the $m$-th propagation path. The term $v(t)$ represents AWGN with a mean of zero and a variance of $\sigma^2$.  The received signal is then sampled at times $t=(\kappa n+i)T_s$, where $\kappa$ and $T_s$ denote the number of samples per symbol and sampling period, respectively, and $i\in\{0,1,...,\kappa-1\}$. The sampling procedure is as follows
\begin{equation}
    r_n^i\triangleq r(t)\vert_{t=(\kappa n+i)T_s}=\sum_{m=0}^{L_c-1}h_mx_{n-m}^i+v_n^i,
\end{equation}
where
\begin{align}
    x_{n-m}^i &\triangleq x(t-mT)\vert_{t=(\kappa n+i)T_s},
    \\
    v_n^i &\triangleq v(t)\vert_{t=(\kappa n+i)T_s}.
\end{align}
Based on \eqref{equ:6}, we can derive the discrete representation of the tilted phase CPM signal as
\begin{equation}
        y_n^i=x_n^ie^{j\psi_n^i},
\end{equation}
where $\psi_n^i\triangleq\pi h(M-1)(\kappa n+i)/\kappa$. We shall define the following vectors
\begin{equation}
    \begin{split}
          \mathbf{r}_n &=[r_n^0,r_n^1,...,r_n^{\kappa-1}]^T, 
          \\
          \mathbf{x}_n &=[x_n^0,x_n^1,...,x_n^{\kappa-1}]^T, 
          \\
          \mathbf{y}_n &=[y_n^0,y_n^1,...,y_n^{\kappa-1}]^T,
          \\
          \mathbf{v}_n &=[v_n^0,v_n^1,...,v_n^{\kappa-1}]^T.
    \end{split}
\end{equation}\par
After removing the CP, the received signal can be represented in matrix form as follows
\begin{equation}
    \mathbf{r}=\mathbf{H}\mathbf{x}+\mathbf{v},
    \label{equ:20}
\end{equation}
where $\mathbf{H}$ is a $\kappa\tilde{N}\times\kappa\tilde{N}$ dimensional cyclic matrix with the first column $\mathbf{h}_0=[h_0,...,h_{L_c-1},\mathbf{0}]^T$, $\mathbf{r}=[\mathbf{r}_0,\mathbf{r}_1,...,\mathbf{r}_{\tilde{N}-1}]^T$, $\mathbf{x}=[\mathbf{x}_0,\mathbf{x}_1,...,\mathbf{x}_{\tilde{N}-1}]^T$, and $\mathbf{v}=[\mathbf{v}_0,\mathbf{v}_1,...,\mathbf{v}_{\tilde{N}-1}]^T$.  

\section{Iterative equalization of CPM using massing passing}

\subsection{Probabilistic Model and Factor Graph}
The matrix $\mathbf{H}$ in \eqref{equ:20} is a dense one,  which will lead to a densely connected factor graph, causing divergence of conventional message passing algorithms. UAMP, a variant of AMP, is promising to handle this, which is hence used in this paper. To facilitate the use of UAMP, a unitary transformation of \eqref{equ:20} needs to be performed. The channel matrix $\mathbf{H}$ can be decomposed by singular value decomposition (SVD) as $\mathbf{H}=\mathbf{U\Lambda V}$. Consequently, the unitary transformation of \eqref{equ:20} is given by
\begin{equation}
    \mathbf{z}=\mathbf{U}^H\mathbf{r}=\mathbf{\Phi x} +\mathbf{w},
    \label{equ:21}
\end{equation}
where $\mathbf{\Phi}=\mathbf{U}^H\mathbf{H}=\mathbf{\Lambda V}$, $\mathbf{\Lambda}$ is a diagonal matrix of size $\kappa\tilde{N}\times\kappa\tilde{N}$, and $\mathbf{w}=\mathbf{U}^H\mathbf{v}$ has the same mean and variance as $\mathbf{v}$. However, in the considered problem, $\mathbf{H}$ is a circulant matrix, which can be diagonalized using a discrete Fourier transform (DFT) or inverse DFT (IDFT) matrix. Hence, SVD is no longer needed, and we can simply set $\mathbf{U}=\mathbf{F}^H$ and $\mathbf{V}=\mathbf{F}$, respectively, where $\mathbf{F}$ is a DFT matrix of size $\kappa\tilde{N}\times\kappa\tilde{N}$. Furthermore, $\mathbf{\Lambda}$ can easily be obtained by DFT with $\mathbf{\Lambda}=\text{diag}(\mathbf{Fh}_0)$. With this setting, \eqref{equ:21} can be regarded as the frequency domain form of \eqref{equ:20}.\par
\begin{figure*}
    \centering
    \includegraphics[width=0.75\linewidth]{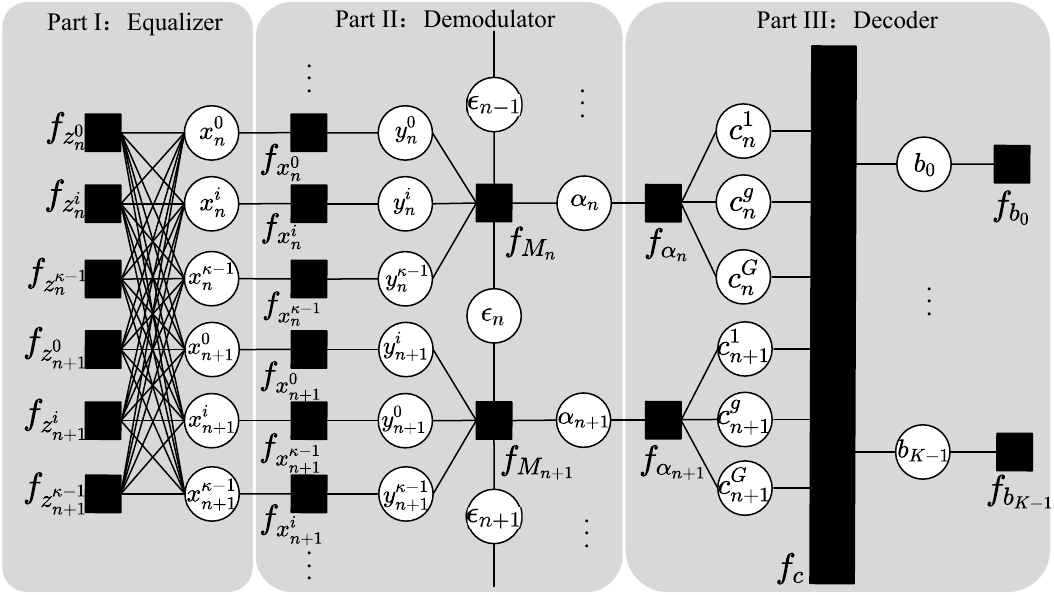}
    \caption{Factor graph representation of (23).}
    \label{fig:2}
\end{figure*}

The received signal after unitary transform during the interval $[nT, (n+1)T]$ can be represented as
\begin{equation}
    \mathbf{z}_n=\mathbf{\Phi}_n\mathbf{x}+\mathbf{w}_n,
\end{equation}
where $\mathbf{z}_n=[z_n^0,z_n^1,...,z_n^{\kappa-1}]^T$, $\mathbf{\Phi}_n=[\mathbf{\Phi}_n^0;\mathbf{\Phi}_n^1;...;\mathbf{\Phi}_n^{\kappa-1}]$, and $\mathbf{w}_n=[w_n^0,w_n^1,...,w_n^{\kappa-1}]^T$. Then, the joint distribution of all variables $\mathbf{b},\mathbf{c},\boldsymbol{\alpha},\mathbf{x},\mathbf{y}$ and $\boldsymbol{\epsilon}$ given by the observation $\mathbf{z}$ can be factorized as
\begin{equation}
    \begin{split}
        &p(\mathbf{b},\mathbf{c},\boldsymbol{\alpha},\mathbf{x},\mathbf{y},\boldsymbol{\epsilon}\vert\mathbf{z})  \\ 
        &\propto\prod_{n,i}f_{z_n^i}(z_n^i,\mathbf{x})f_{x_n^i}(x_n^i,y_n^i)\\ 
        &\quad\times \prod_nf_{\mathbf{y}_n}(\mathbf{y}_n,\epsilon_n,\alpha_n)f_{\epsilon_n}(\epsilon_n,\epsilon_{n-1},\alpha_n)\\
        &\quad\times f_{\alpha_n}(\alpha_n,\mathbf{c}_n)f_c(\mathbf{b},\mathbf{c})\prod_kf_{b_k}(b_k) \\
        &\triangleq\prod_{n,i}f_{z_n^i}(z_n^i,\mathbf{x})f_{x_n^i}(x_n^i,y_n^i)\\
        &\quad\times\prod_nf_{M_n}(\mathbf{y}_n,\epsilon_n,\epsilon_{n-1},\alpha_n) \\
        &\quad\times f_{\alpha_n}(\alpha_n,\mathbf{c}_n)f_c(\mathbf{b},\mathbf{c})\prod_kf_{b_k}(b_k), \\
    \end{split}
    \label{equ:23}
\end{equation}
where 
\begin{equation}
    f_{M_n}(\mathbf{y}_n,\epsilon_n,\epsilon_{n-1},\alpha_n)=f_{\mathbf{y}_n}(\mathbf{y}_n,\epsilon_n,\alpha_n)f_{\epsilon_n}(\epsilon_n,\epsilon_{n-1},\alpha_n)
\end{equation}
represents the aggregation of factor nodes for $f_{\mathbf{y}_n}(\mathbf{y}_n,\epsilon_n,\alpha_n)$ and $f_{\epsilon_n}(\epsilon_n,\epsilon_{n-1},\alpha_n)$, which is designed to prevent loops within the factor graph of  the CPM demodulation process. The probabilistic functions associated with the remaining factors are detailed in Table \ref{tab:I}. The factor graph corresponding to \eqref{equ:23} is illustrated in Fig. \ref{fig:2}. Our aim is to find the (approximate) posteriors $\{p(b_k|\mathbf{z})\}$ through message passing, based on which hard decision on the information bits can be made. \par
\begin{table}[t]
  \centering
  \caption{FACTORS AND DISTRIBUTIONS IN \eqref{equ:23}}
  \label{tab:I}
  \begin{tabular*}{0.9\linewidth}{@{}ccc@{}}
    \toprule
    \makebox[0.12\linewidth][c]{Factor}& \makebox[0.3\linewidth][c]{Distribution}& \makebox[0.3\linewidth][c]{Function}\\
    \midrule
    $f_{z_n^i}$&$p(z_n^i|\mathbf{x})$ & $\mathcal{CN}(z_n^i;\mathbf{\Phi}_n^i\mathbf{x},\sigma^2)$\\
    $f_{x_n^i}$& $p(x_n^i|y_n^i)$& $\delta(x_n^i-y_n^ie^{-j\psi_n^i}) $\\
    $f_{\mathbf{y}_n}$&$p(\mathbf{y}_n|\epsilon_n,\alpha_n)$& $\delta(\mathbf{y}_n-\boldsymbol{\chi}_l)$\\
    $f_{\epsilon_n}$&$p(\epsilon_n|\epsilon_{n-1},\alpha_n)$& state transition\\
    $f_{\alpha_n}$&$p(\alpha_n|\mathbf{c}_n)$& mapping constraints\\
    $f_c$&$p(\mathbf{c}|\mathbf{b})$& coding constraints\\
    $f_{b_k}$& $p(b_k)$& prior distribution of $b_k$\\
    \bottomrule
  \end{tabular*}
\end{table}
As illustrated in Fig. \ref{fig:2}, the nodes within the equalizer section exhibit a high degree of connectivity. In this case, the UAMP algorithm is capable of reducing computational complexity and accelerating convergence, while exhibiting conceptional robustness against a general matrix $\mathbf{H}$ \cite{Ref33,Ref37,Ref38} . The implementation of the demodulator and decoder employ the BP rule \cite{Ref39}, which is particularly effective at handling discrete variables. Thus, the factor nodes are divided into two disjoint subsets: one for equalization (UAMP) and the other for demodulation and decoding (BP)
\begin{align}
    \mathcal{A}_{\text{UAMP}}&=\{f_{z_n^i}\}, \nonumber
    \\
    \mathcal{A}_{\text{BP}}&=\{f_{x_n^i}\}\cup\{f_{M_n}\}\cup\{f_{\alpha_n}\}\{f_c\}\cup\{f_{b_k}\},
\end{align}
where $\mathcal{A_\text{UAMP}}$ and $\mathcal{A_\text{BP}}$ denote the respective sets of factor nodes associated with the UAMP and BP parts, respectively. We denote by $n_{A\to B}(x)$ the message from a variable node $A$ to a factor node $B$, and by $m_{B\to A}(x)$ the message from a factor node $B$ to a variable node $A$.
\subsection{Iterative receiver design}
\subsubsection{Part I-Equalizer} 
The equalizer part, as shown in Fig. \ref{fig:2}, primarily employs the UAMP algorithm to estimate the discrete data symbols in the vector $\mathbf{x}$. The mean and variance of these symbols can be calculated using the functions $g_x(\mathbf{q},\boldsymbol{\tau}_q)$ and $g'_x(\mathbf{q},\boldsymbol{\tau}_q)$, respectively.  The function $g_x(\mathbf{q},\boldsymbol{\tau}_q)$ yields a column vector, with the $(n\kappa+i)$-th component designated as $[g_x(\mathbf{q},\boldsymbol{\tau}_q)]_n^i$, calculated by 
\begin{equation}
    [g_x(\mathbf{q},\boldsymbol{\tau}_q)]_n^i=\frac{\int x_n^ip(x_n^i)\mathcal{CN}(x_n^i;q_n^i,\tau_{q_n^i})dx_n^i}{\int p(x_n^i)\mathcal{CN}(x_n^i;q_n^i,\tau_{q_n^i})dx_n^i},
\label{equ:26}
\end{equation}
where $p(x_n^i)$ represents the prior distribution of $x_n^i$, $\mathcal{CN}(x_n^i;q_n^i,\tau_{q_n^i})$ denotes a complex normal distribution with mean $q_n^i$ and variance $\tau_{q_n^i}$. Eq. \eqref{equ:26} represents the MMSE estimator of $x_n^i$ inferred from the observation model, which is given by
\begin{equation}
    q_n^i=x_n^i+\nu_n^i,
\end{equation}
where $\nu_n^i$ represents the AWGN with zero mean and variance $\tau_{q_n^i}$. The function $g'_x(\mathbf{q},\boldsymbol{\tau}_q)$ furnishes another column vector, with its $(n\kappa+i)$-th entry being  $[g'_x(\mathbf{q},\boldsymbol{\tau}_q)]_n^i$, differentiated with respect to $q_{n,i}$. Thus the estimation of $x_n^i$ can be divided into two parts: 
\begin{enumerate}[i)]
    \item The prior distribution of $x_n^i$, which is represented the input message $m_{f_{x_n^i}\to x_n^i}(x_n^i)$ of the equalizer and subsequently defined in \eqref{equ:52};
    \item The observation information of $x_n^i\sim\mathcal{CN}(x_n^i;q_n^i,\tau_{q_n^i})$, which is the output  of the equalizer.
\end{enumerate}

The values of $q_n^i$ and $\tau_{q_n^i}$ can be obtained via the following steps \cite{Ref33}: First, generate two vectors
\begin{equation}
    \boldsymbol{\tau}_p=|\mathbf{\Phi}|^2\boldsymbol{\tau}_x,
    \label{equ:28}
\end{equation}
\begin{equation}
    \mathbf{p}=\mathbf{\Phi \hat{x}}-\boldsymbol{\tau}_p \cdot\mathbf{s},
    \label{equ:29}
\end{equation}
where $\mathbf{s}$ is a vector obtained from the previous iteration. Then the intermediate vectors, namely  $\boldsymbol{\tau}_s$ and $\mathbf{s}$, are updated as follows
\begin{equation}
    \boldsymbol{\tau}_s=\mathbf{1}./(\boldsymbol{\tau}_p+\sigma^2\mathbf{1}),
    \label{equ:30}
\end{equation}
\begin{equation}
    \mathbf{s}=\boldsymbol{\tau}_s\cdot(\mathbf{z}-\mathbf{p}).
    \label{equ:31}
\end{equation} 
Finally, the vectors $\boldsymbol{\tau}_q$ and $\mathbf{q}$ are calculated using the following formulas 
\begin{equation}
    \mathbf{1}./\boldsymbol{\tau}_q=|\mathbf{\Phi}^H|^2\boldsymbol{\tau}_s,
    \label{equ:32}
\end{equation}
\begin{equation}
   \mathbf{q}=\hat{\mathbf{x}}+\boldsymbol{\tau}_q\cdot\mathbf{\Phi}^H\mathbf{s}.
   \label{equ:33}
\end{equation}
\par

 Given the input message $m_{f_{x_n^i}\to x_n^i}(x_n^i)$ of the equalizer, the belief $b(x_n^i)$ corresponding to the variable $x_n^i$ is derived as follows
\begin{equation}
    \begin{split}
        b(x_n^i)&=n_{x_n^i\to f_{x_n^i}}(x_n^i)m_{f_{x_n^i}\to x_n^i}(x_n^i) \\
        &\propto\mathcal{CN}(x_n^i;\hat x_n^i,\hat\tau_{x_n^i}), 
    \end{split}
\end{equation}
where
\begin{equation}
    \hat x_n^i=e^{-j\psi_n^i}\sum_{l=0}^{PM^L-1}\chi_l^i P(y_n^i=\chi_l^i),
    \label{equ:35}
\end{equation}
\begin{equation}
    \hat\tau_{x_n^i}=\sum_{l=0}^{PM^L-1}|\hat x_n^ie^{j\psi_n^i}-\chi_l^i|^2P(y_n^i=\chi_l^i),
    \label{equ:36}
\end{equation}
where 
\begin{equation}
    P(y_n^i=\chi_l^i)=\frac{P_l(y_n^i=\chi_l^i)}{\sum_{l'=0}^{PM^L-1}P_{l'}(y_n^i=\chi_{l'}^i)},
    \label{equ:37}
\end{equation}
denote the normalized probability and
\begin{equation}
   P_l(y_n^i=\chi_l^i)=P^a(y_n^i=\chi_l^i)\cdot\text{exp}\left(-\frac{|\chi_l^i-q_n^ie^{j\psi_n^i}|^2}{\tau_{q_n^i}}\right),
   \label{equ:38}
\end{equation}
where $P^a(y_n^i=\chi_l^i)$ represents the prior probability of $y_n^i=\chi_l^i$ and is defined later in \eqref{equ:50}.
\par

In the scalar stepsize UAMP, it is possible to employ the mean value of  $\hat\tau_{x_n^i}$, denoted as $\tau_x$, in place of  $\hat\tau_{x_n^i}$ itself, thus circumventing the need for matrix-vector multiplication in \eqref{equ:28} and  \eqref{equ:32} \cite{Ref33}. The $\tau_x$ is calculated by
\begin{equation}
    \tau_x=\frac{1}{\kappa\tilde{N}}\sum_{n=0}^{\tilde{N}-1}\sum_{i=0}^{\kappa-1}\hat\tau_{x_n^i}.
    \label{equ:39}
\end{equation}
 Consequently, \eqref{equ:28} and \eqref{equ:32}  can be  simplified to
\begin{equation}
    \boldsymbol{\tau}_p=\tau_x\boldsymbol{\lambda},
    \label{equ:40}
\end{equation}
\begin{equation}
     1/\tau_q=(1/\tilde{N})\boldsymbol{\lambda}^H\boldsymbol{\tau}_s,
     \label{equ:41}
\end{equation}
where $\boldsymbol{\lambda}=\mathbf{\Lambda\Lambda}^H\mathbf{1}$. 

\subsubsection{Part II-Demodulator}
The message from the factor $f_{x_n^i}$ to the variable node $y_n^i$ can be calculated by the BP rule, which is derived from the forward message $n_{x_n^i\to f_{x_n^i}}(x_n^i)$ passed from the UAMP. The message $m_{f_{x_n^i\to y_n^i}}(y_n^i)$ essentially represents the transformation from the equalized CPM data to the tilted phase CPM and can be expressed as
\begin{align}
\label{equ:42}
    m_{f_{x_n^i}\to y_n^i}(y_n^i)&=\sum_{x_n^i} f_{x_n^i}(x_n^i)\cdot n_{x_n^i\to f_{x_n^i}}(x_n^i)  \nonumber \\
    &\propto\mathcal{CN}(y_n^i;q_n^ie^{j\psi_n^i},\tau_q).
\end{align}
\par

In order to facilitate the calculation of messages in the demodulator of CPM, we propose to denote all messages transmitted from the variable nodes $\mathbf{y_n}$ to the specific factor node $f_{M_n}$ at the time instance $n$ as $n_{\mathbf{y}_n\to f_{M_n}}(\mathbf y_n)$, while those in the reverse direction are denoted as $m_{f_{M_n}\to\mathbf y_n}(\mathbf y_n)$. Given the above notations, we can use the BP rule to compute the downward message for the demodulator part of the CPM as 
\begin{equation}
    \begin{split}
        m&_{f_{M_n}\to\epsilon_n}(\epsilon_n)\\     
        =&\sum_{\mathbf{y}_n}\sum_{\epsilon_{n-1}}\sum_{\alpha_n}f_{\mathbf{y}_n}(\mathbf{y}_n,\epsilon_n,\alpha_n)f_{\epsilon_n}(\epsilon_n,\epsilon_{n-1},\alpha_n) \\
        &\times n_{\mathbf{y}_n\to f_{M_n}}(\mathbf{y}_n)n_{\epsilon_{n-1}\to f_{M_n}}(\epsilon_{n-1})n_{\alpha_n\to f_{M_n}}(\alpha_n)\\
        =&\sum_{\epsilon_{n-1}}\sum_{\alpha_n}p(\mathbf{q}_ne^{j\boldsymbol{\psi}_n}|\boldsymbol{\chi}_l,\tau_q)m_{f_{M_{n-1}}\to\epsilon_{n-1}}(\epsilon_{n-1})\\
        &\times n_{\alpha_n\to f_{M_n}}(\alpha_n).
    \end{split}
    \label{equ:43}
\end{equation}
The above equation is similar to the forward recursion formula of the BCJR algorithm, with the initial condition given by 
\begin{equation}
    m_{f_{M_{-1}}\to\epsilon_{-1}}(\epsilon_{-1})=p(\epsilon_{-1}),
\end{equation} 
where $p(\epsilon_{-1})$ represents the prior distribution of the state $\epsilon_{-1}$ with the time index $-1$, which is  assumed to be known. Additionally, $p(\mathbf{q}_ne^{j\boldsymbol{\psi}_n}|\boldsymbol{\chi}_l,\tau_q)$, which serves as the branching metric in this context, is calculated by
\begin{equation}
      p(\mathbf{q}_ne^{j\boldsymbol{\psi}_n}|\boldsymbol{\chi}_l,\tau_q)\propto\text{exp}\left[{-\sum_{i=0}^{\kappa-1}(q_n^ie^{j\psi_n^i}-\chi_l^i)^2/\tau_q}\right].       
     \label{equ:45}
\end{equation}
$n_{\alpha_n\to f_{M_n}}(\alpha_n)$ contains the prior information of $\alpha_n$ and is updated by the output of the channel decoding. Finally, it is crucial to highlight that within the context of the aforementioned equation, the variables  $\epsilon_n$, $\epsilon_{n-1}$ and $\alpha_n$ are subject to the constraint $f_{\epsilon_n}(\epsilon_n,\epsilon_{n-1},\alpha_n)=1$, which indicates a deterministic relationship between these variables within the factor $f_{\epsilon_n}$ and is implied in the expression. \par
Similarly, the upward message for the demodulator part of the CPM can be derived, which corresponds to the backward recurrence formula of BCJR
\begin{equation}
    \begin{split}
        m&_{f_{M_n}\to\epsilon_{n-1}}(\epsilon_{n-1})\\     
        =&\sum_{\mathbf{y}_n}\sum_{\epsilon_n}\sum_{\alpha_n}f_{\mathbf{y}_n}(\mathbf{y}_n,\epsilon_n,\alpha_n)f_{\epsilon_n}(\epsilon_n,\epsilon_{n-1},\alpha_n)\\
        &\times n_{\mathbf{y}_n\to f_{M_n}}(\mathbf{y}_n)n_{\epsilon_n\to f_{M_n}}(\epsilon_n)n_{\alpha_n\to f_{M_n}}(\alpha_n)\\
        =&\sum_{\epsilon_n}\sum_{\alpha_n}p(\mathbf{q}_ne^{j\boldsymbol{\psi}_n}|\boldsymbol{\chi}_l,\tau_q) m_{f_{M_{n+1}}\to\epsilon_n}(\epsilon_n)\\
        &\times n_{\alpha_n\to f_{M_n}}(\alpha_n),
    \end{split}
    \label{equ:46}
\end{equation}
with the initial condition provided by 
\begin{equation}
    m_{f_{M_{\tilde{N}}}\to \epsilon_{\tilde{N}-1}}(\epsilon_{\tilde{N}-1})=\frac{1}{PM^{L-1}}.
\end{equation}
\par
Subsequently, the right output message for the demodulator part of the CPM can be calculated as follows
\begin{equation}
    \begin{split}
        m&_{f_{M_n}\to\alpha_n}(\alpha_n)\\
        =&\sum_{\mathbf{y}_n}\sum_{\epsilon_n}\sum_{\epsilon_{n-1}}f_{\mathbf{y}_n}(\mathbf{y}_n,\epsilon_n,\alpha_n)f_{\epsilon_n}(\epsilon_n,\epsilon_{n-1},\alpha_n)\\
        &\times n_{\mathbf{y}_n\to f_{M_n}}(\mathbf{y}_n)n_{\epsilon_{n-1}\to f_{M_n}}(\epsilon_{n-1})n_{\epsilon_n\to f_{M_n}}(\epsilon_n)\\
        =&\sum_{s_n}p(\mathbf{q}_ne^{j\boldsymbol{\psi}_n}|\boldsymbol{\chi}_l,\tau_q)m_{f_{M_{n-1}}\to\epsilon_{n-1}}(\epsilon_{n-1})\\
        &\times m_{f_{M_{n+1}}\to\epsilon_n}(\epsilon_n),
    \end{split}
    \label{equ:48}
\end{equation}
and the left output message for the demodulator part of the CPM can be calculated as follows
\begin{equation}
    \begin{split}
        m&_{f_{M_n}\to\mathbf{y}_n}(\mathbf{y}_n) \\
        =&\sum_{\epsilon_n}\sum_{\epsilon_{n-1}}\sum_{\alpha_n}f_{\mathbf{y}_n}(\mathbf{y}_n,\epsilon_n,\alpha_n)f_{\epsilon_n}(\epsilon_n,\epsilon_{n-1},\alpha_n) \\
        &\times n_{\epsilon_{n-1}\to f_{M_n}}(\epsilon_{n-1})n_{\epsilon_n\to f_{M_n}}(\epsilon_n)n_{\alpha_n\to f_{M_n}}(\alpha_n) \\
        =&\sum_{l=0}^{PM^L-1}\boldsymbol{\chi}_lP^a(\mathbf{y}_n=\boldsymbol{\chi}_l),
    \end{split}
    \label{equ:49}
\end{equation}
where
\begin{equation}
   \begin{split}
         P^a(\mathbf{y}_n=\boldsymbol{\chi}_l)\triangleq & ~m_{f_{M_{n-1}}\to\epsilon_{n-1}}(\epsilon_{n-1})m_{f_{M_{n+1}}\to \epsilon_n}(\epsilon_n)\\
         &\times n_{\alpha_n\to f_{M_n}}(\alpha_n),  
   \end{split}
   \label{equ:50}
\end{equation}

The above expression is not strict because the probability $P^a(\mathbf y_n=\mathbf\chi_l)$ should be normalized. However, this expression is only an intermediate variable and does not affect the final result, and the normalized probability is given in \eqref{equ:37}.  \par
In \eqref{equ:49}, the value of the $i$-th sampling point is given by 
\begin{equation}
        m_{f_{M_n}\to y_n^i}(y_n^i)=\sum_{l=0}^{PM^L-1}\chi_l^iP^a(y_n^i=\chi_l^i).
\end{equation}

Consequently, the message from $f_{x_n^i}$ to $x_n^i$, which is the input of UAMP, can be calculated as
\begin{align}
\label{equ:52}
    m_{f_{x_n^i}\to x_n^i}(x_n^i)&=\sum_{y_n^i}n_{y_n^i\to f_{x_n^i}}(y_n^i) f_{x_n^i}(x_n^i) \nonumber\\
    &= e^{-j\psi_n^i}\sum_{l=0}^{PM^L-1}\chi_l^iP^a(y_n^i=\chi_l^i).
\end{align} \par
It is essential to emphasize that, as described in \eqref{equ:38}, \eqref{equ:43}, \eqref{equ:45}, \eqref{equ:46}, \eqref{equ:48}, and \eqref{equ:50}, the operation occurs within the linear-domain. However, for practical feasibility, it will be implemented in the log-domain. 

\subsubsection{Part III-Decoder}
The decoder is implemented using the log-domain BCJR algorithm. The extrinsic information for each code bit $c_n^g$ is given by 
\begin{equation}
     L^e(c_n^g)=\text{ln}\frac{P(c_n^g=0|\mathbf z)}{P(c_n^g=1|\mathbf z)}-L^a(c_n^g),
     \label{equ:53}
\end{equation}
where $L^a(c_n^g)$ represents the output extrinsic log-likelihood ratio (LLR) of the decoder. However, for the turbo equalization of CPM signals, the symbol values of CPM obtained through extrinsic information rapidly converge to zero, preventing any occurrence of SIC. The convergence is substantiated by the Laurent decomposition method as detailed in \cite[Appendix]{Ref22}. Therefore, in the UAMP algorithm, we utilize bit information rather than extrinsic information as the prior information returned by the decoder. Furthermore, \eqref{equ:53} can be concisely represented by the following formula \cite{Ref40}
\begin{equation}
    L^e(c_n^g)=\text{ln}\frac{\underset{\beta_j\in\mathcal{A}_{g,0}}{\sum}m_{f_{M_n}\to\alpha_n}(\alpha_n)\underset{g'\neq g}{\prod}P(c_n^{g'}=\beta_j^{g'})}{\underset{\beta_j\in\mathcal{A}_{g,1}}{\sum}m_{f_{M_n}\to\alpha_n}(\alpha_n)\underset{g'\neq g}{\prod}P(c_n^{g'}=\beta_j^{g'})},
    \label{equ:54}
\end{equation}
where $\beta_j$ represents a member of the set of possible values for $\alpha_n$. This set is defined as follows: $\alpha_n\in\mathcal{A}=[\beta_0,\beta_1,...,\beta_{M-1}]$. Each element $\beta_j$ of the set $\mathcal{A}$ is associated with a length-$G$ binary sequence, which is represented by the notation $[\beta_j^0, \beta_j^1, ..., \beta_j^{G-1}]$. $\mathcal{A}_{g,0}$ and $\mathcal{A}_{g,1}$ represent the subset of all $\beta_j$ with $\beta_j^g=0$ and $\beta_j^g=1$, respectively. 

\subsection{Message passing scheduling}
Fig. \ref{fig:2} comprises three inner iterations: the inner iteration of the equalizer, the inner iteration between the equalizer and the demodulator, and the inner iteration between the demodulator and the decoder. However, we only keep the inner iterations between the demodulator and the decoder. The reason is that the robust error correction capabilities of channel coding enhance the prior information of the equalizer through multiple channel decoding iterations. Meanwhile, the other two inner iterations are merged into the outer iteration to reduce unnecessary iterations. Algorithm \ref{algorithmic1} provides an overview of the computation order for the proposed message passing algorithm. First, the UAMP algorithm is used to calculate the equalizer output  (Part I with lines $2$-$5$). This output is then passed to the demodulator part to compute the upward and downward messages, as well as the forward output of the demodulator (as shown in lines $7$-$10$ of Part II forward). Then, the BCJR algorithm updates the prior information of the transmitted symbols (Part III with lines $11$-$13$). Finally, the backward output of the CPM demodulator (Part II backward with lines $15$-$16$) is calculated and used as input to the equalizer for the next iteration. This process is repeated until the predefined maximum number of iterations is reached. The entire iteration process is denoted as the outer iteration, with the iteration number denoted as $n_o$. Additionally, the inner iteration, which consists of demodulator forward and decoder parts as outlined in lines $6$-$14$, is identified by the iteration number $n_i$. \par

\begin{algorithm}[t]
     \caption{Iterative equalization of CPM with UAMP.}
     \textbf{Unitary transform}: $\mathbf{z}=\mathbf{U}^H\mathbf{r}=\mathbf{\Phi x}+\mathbf{w}$, where $\mathbf\Phi=\mathbf{U}^H\mathbf{H}=\mathbf{\Lambda V}$, $\mathbf{\Lambda}=\text{Diag}(\mathbf{Fh}_0)$, $\mathbf{w}=\mathbf{Fv}$, $\mathbf{U}=\mathbf{F}^H$, and $\mathbf{V}=\mathbf{F}$.\\
     Define vector $\boldsymbol{\lambda}=\mathbf{\Lambda\Lambda}^H\mathbf{1}$.\\
     \textbf{Initialize:} $\mathbf{\hat{x}}=\mathbf{0}$, $\tau_x=1$, $\mathbf{s}=\mathbf{0}$, $m_{f_{M_{-1}}\to s_{-1}}(s_{-1})=p(s_{-1})$, $ m_{f_{M_{\tilde N}}\to \epsilon_{\tilde N-1}}(\epsilon_{\tilde N-1})=\frac{1}{PM^{L-1}}$, $L^a(c_{n,g})=\text{ln}(0.5)$. 
     \begin{algorithmic}[1] 
     \For{$it\_out=1:n_o$} 
           \Statex \textbf{\% Part I - Equalizer}
           \State $\forall n,i$  update  $\hat{x}_n^i$ and $\tau_x$ with \eqref{equ:35} and \eqref{equ:39};
           \State  $\forall n,i$  update  $\tau_{p_n^i}$ and  $p_n^i$ with \eqref{equ:40} and \eqref{equ:29};
           \State  $\forall n,i$  update  $\tau_{s_n^i}$ and $s_n^i$ with \eqref{equ:30} and \eqref{equ:31};
           \State $\forall n,i$  update  $1/\tau_q$ and $q_n^i$ with \eqref{equ:41} and \eqref{equ:33};
           \For{$it\_inner=1:n_i$}
                  \Statex \hspace{0.9cm} \textbf{\% Part II - Demodulator - forward}
                  \State $\forall n,i$  update  $ m_{f_{x_n^i}\to y_n^i}$ with \eqref{equ:42};
                  \State $\forall n$  update $m_{f_{M_n}\to\epsilon_n}(\epsilon_n)$ with \eqref{equ:43};
                  \State $\forall n$  update $m_{f_{M_n}\to\epsilon_{n-1}}(\epsilon_{n-1})$ with \eqref{equ:46};
                  \State $\forall n$  update $m_{f_{M_n}\to\alpha_n}(\alpha_n)$ with \eqref{equ:48};
                  \Statex \hspace{0.9cm} \textbf{\% Part III - Decoder}
                  \State $\forall n,g$  update $L^e(c_n^g)$ with \eqref{equ:54}; 
                  \State  $\forall n,g$  update $L^a(c_n^g)$ with BCJR;
                  \State $\forall n$ update $n_{\alpha_n\to f_{M_n}}(\alpha_n)$ by $L^a(c_n^g)$;
           \EndFor
           \Statex \textbf{\% Part II - Demodulator - backward}
           \State $\forall n$ update $m_{f_{M_n}\to \mathbf{y}_n}(\mathbf y_n)$ with \eqref{equ:49};
           \State $\forall n,i$  update  $m_{f_{x_n^i}\to x_n^i}(x_n^i)$ with \eqref{equ:52}.
     \EndFor
     \end{algorithmic}
     \label{algorithmic1}
\end{algorithm}

\subsection{Complexity comparison}
The proposed algorithm consists of three components: the equalizer, the CPM demodulator, and the channel decoder. The complexity of the UAMP algorithm is primarily determined by calculating the mean and variance of the vector $\mathbf x$ in line $2$ and performing the matrix-vector product operations in lines $3$ and $5$. The computational complexity of the variables $\hat{x}_n^i$ and $\tau_x$ in line $2$ is $O(\kappa\tilde{N}PM^L)$. For lines $3$ and $5$, the incorporation of CP allows for the replacement of matrix-vector products with DFT and IDFT operations, resulting in a complexity of $O(\kappa\tilde{N}\text{log}_2\kappa\tilde{N})$. Therefore, the complexity of the equalizer part is $O(\kappa\tilde{N}\text{log}_2\kappa\tilde{N})+O(\kappa\tilde{N}PM^L)$. The complexity of the CPM demodulator, as quantified in lines $7$-$9$, is $O(\kappa\tilde{N}PM^L)$. The complexity associated with the BCJR algorithm for channel decoder is $O(K2^{L_c+1})$. The overall complexity of the Algorithm \ref{algorithmic1} is presented in Table \ref{tab:II}, where $L_f$ represents the length of the time domain equalization (TDE) filter. 

As shown in Table~\ref{tab:II}, the proposed equalization algorithm has equivalent computational complexity to the MMSE/SIC-FDE algorithm. Although the two algorithms appear to be computationally similar, their computational loads are not identical. Specifically, within the MMSE/SIC algorithm, the normalization in \eqref{equ:37} adopts $P^a(y_n^i=\chi_l^i)$ rather than $P_l(y_n^i=\chi_l^i)$, resulting in a slight complexity reduction compared to the UAMP algorithm. This leads to an unavoidable consequence that affects the performance of the MMSE approach. Furthermore, it should be noted that preprocessing complexities, such as filter computation in MMSE/SIC-FDE and matrix inversion in MMSE/SIC-TED, are not included in this analysis. These processes can be executed offline, thereby excluding them from iterative computational considerations. Consequently, Table \ref{tab:II} only reflects the complexity relevant to the iterative computations.

\begin{table*}
  \centering
  \caption{COMPARISON OF ALGORITHM COMPLEXITY}
  \label{tab:II}
  \begin{tabular}{lccc}
    \toprule
    \makecell[c]{\textbf{Algorithm}}& \textbf{Complexity}\\
    \midrule
    Proposed algorithm&$O(n_o\kappa\tilde{N}\text{log}_2\kappa\tilde{N})+O(n_o\kappa\tilde{N}PM^L)+O(n_on_i\tilde{N}PM^L)+O(n_on_iK2^{L_c+1})$ \\
    MMSE/SIC-FDE \cite{Ref23}& $O(n_o\kappa\tilde{N}\text{log}_2\kappa\tilde{N})+O(n_o\kappa\tilde{N}PM^L)+O(n_on_i\tilde{N}PM^L)+O(n_on_iK2^{L_c+1})$\\
    MMSE/SIC-TDE \cite{Ref22}& $O(n_o\kappa NL_f)+O(n_o\kappa NPM^L)+O(n_on_iNPM^L)+O(n_on_iK2^{L_c+1})$\\
    \bottomrule
  \end{tabular}
\end{table*}

\section{Simulation}
In this section, we evaluate the performance of the proposed algorithm. We adopt the CPM signal with modulation order $M=4$, correlation length $L=2$, modulation index $h=1/3$, and a raised cosine-shaped pulse. The simulation uses two different channel models: the six-tap typical urban (TU-6) channel and the Proakis’ C channel, which has a higher degree of ISI. Each frame is assumed to experience independent multipath fading, with the channel response remaining constant within a frame. Furthermore, it is assumed that the receiver is perfectly synchronized. The normalized power of the TU-6 channel is given by the vector $[0.189,0.379,0.255,0.090,0.055,0.032]$, with corresponding path delays of $[0, T_s, 2T_s, 8T_s, 12T_s, 25T_s]$. Additionally, the CP length is set to $N_P=13$. For the Proakis’ C channel, the normalized power is $[0.227,0.460,0.688,0.460,0.227]$, with corresponding path delays of $[0, T_s, 2T_s, 3T_s, 4T_s]$. In this instance, the CP length is set to $N_P=2$. The number of sample points per symbol period is set to $\kappa=2$. The channel coding employs a recursive systematic convolutional code with the generating polynomial $[5,7]_8$ and a code rate of $1/2$. Interleaving is performed using pseudo-random techniques. The length of the transmitted data $N$ is set to $508$, while the length for the intrafix is $N_K=2$. Consequently, the length of the sequence after inserting the intrafix is $\tilde{N}=512$. \par

\begin{figure}
    \centering
    \includegraphics[width=1\linewidth]{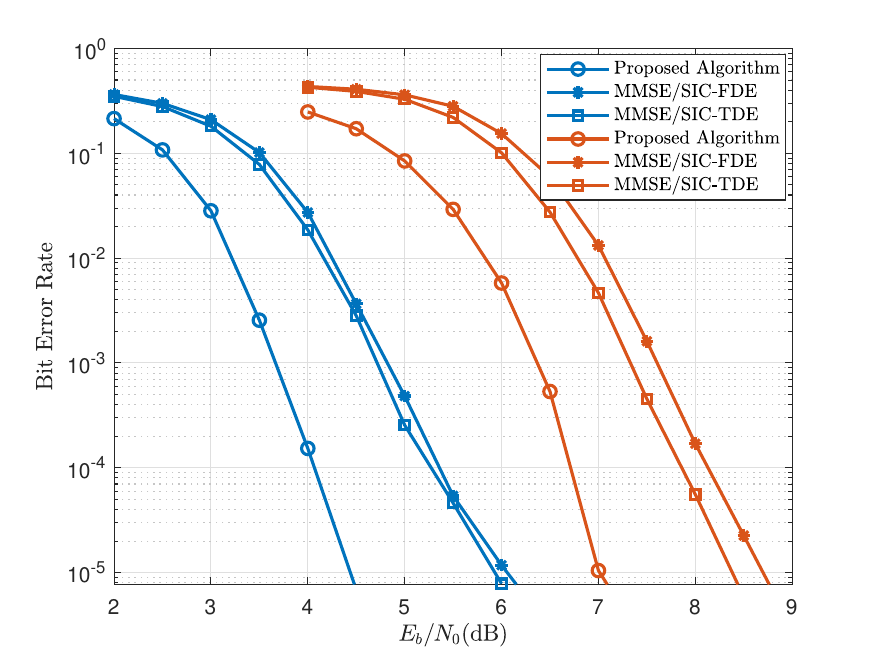}
    \caption{BER performance of the receiver with various equalizers. The blue and orange curves represent the TU-6 and Proakis’ C channels, respectively.}
    \label{fig:3}
\end{figure}

We first simulate the bit error rate (BER) performance of the proposed algorithm with $n_i=1$, which means the algorithm~\ref{algorithmic1} does not utilize inner iteration. Fig. \ref{fig:3} depicts the BER performance of each algorithm after $20$ iterations under the TU-6 and Proakis’ C channels, respectively. As shown in Fig. \ref{fig:3}, the proposed algorithm outperforms the benchmark algorithm regardless of the channel models. In particular, the proposed algorithm shows an approximate performance gain of $1.5$dB and $1.7$dB compared to the traditional MMSE/SIC-based algorithm when the error rate is $10^{-5}$ under the TU-6 and Proakis’ C channels, respectively. 

\begin{figure}
    \centering
    \includegraphics[width=1\linewidth]{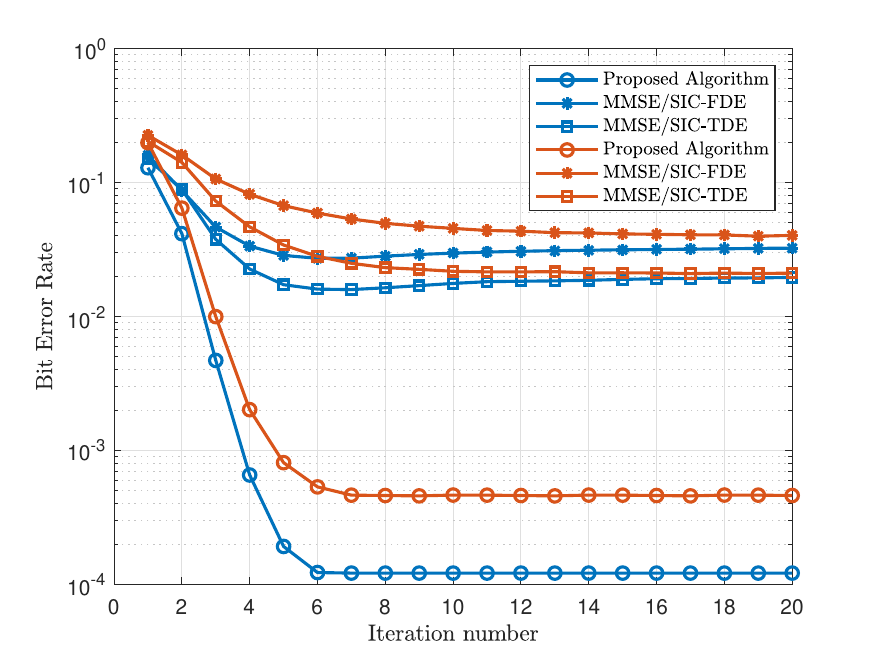}
    \caption{BER performance of the receiver with various numbers of iterations. The blue and orange curves represent the TU-6 channels at $4$dB and Proakis’ C channel at $6.5$dB, respectively.}
    \label{fig:4}
\end{figure}

Fig.~\ref{fig:4} illustrates the convergence speed of the three algorithms in the TU-6 channel and Proakis’ C channel at $E_b/N_0=4$dB and $E_b/N_0=6.5$dB, respectively. As shown in Fig. \ref{fig:4}, under the TU-6 channel, the convergence speed of the proposed algorithm is the same as the MMSE/SIC-based algorithm. The above algorithm achieves convergence after 6 iterations. However, under the Proakis’ C channel with more severe conditions, the proposed algorithm converges after seven iterations, while the MMSE/SIC-based algorithm requires more iterations, and the BER becomes stable after approximately ten  iterations.\par

\begin{figure}[htbp]
    \centering
    \subfloat[$n_i=1$]{\includegraphics[width=0.49\linewidth]{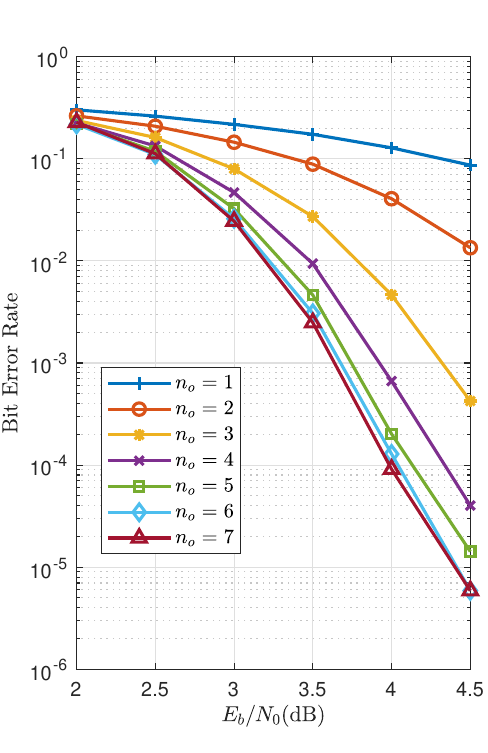}}
        \label{fig:5-1}
    \subfloat[$n_i=2$]{\includegraphics[width=0.49\linewidth]{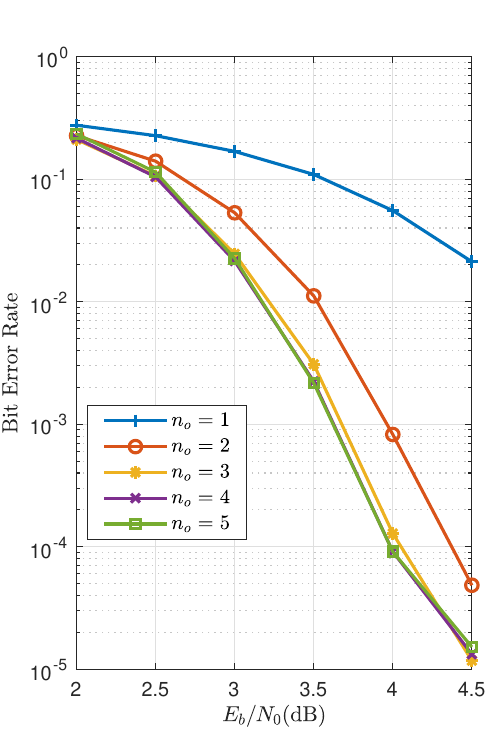}}
        \label{fig:5-2}
    \subfloat[$n_i=3$]{\includegraphics[width=0.49\linewidth]{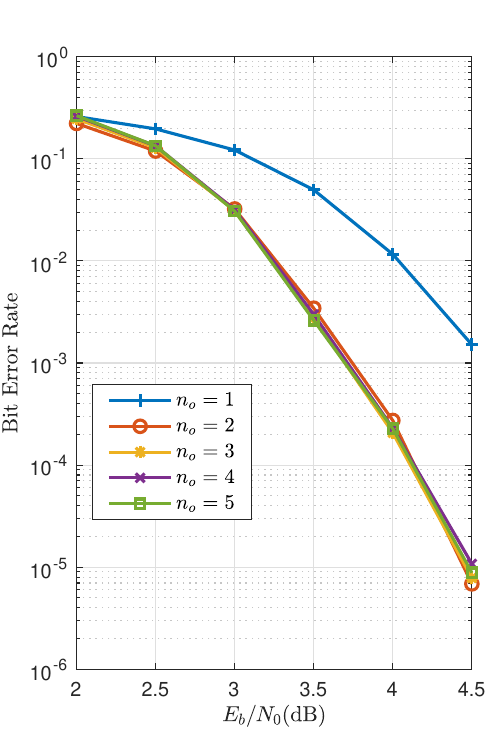}}
        \label{fig:5-3}
    \subfloat[$n_i=4$]{\includegraphics[width=0.49\linewidth]{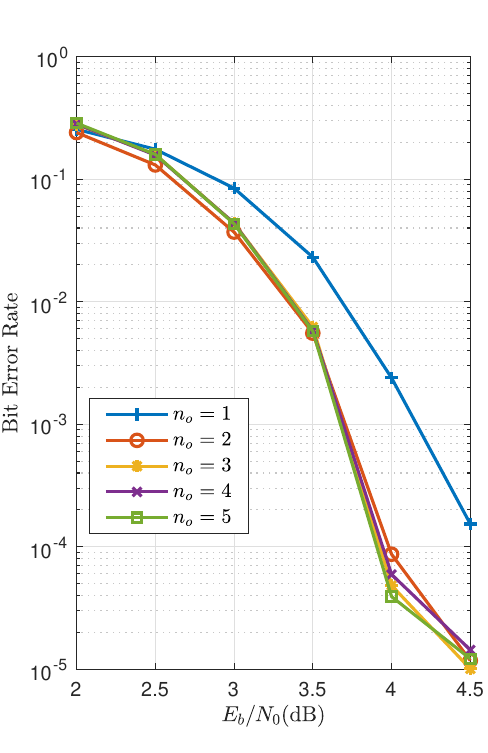}}
        \label{fig:5-4}
    \caption{Convergence performance of the receiver with various numbers of inner iterations under the TU-6 channel.}
    \label{fig:5}
\end{figure}

\begin{figure}
    \centering
    \includegraphics[width=1\linewidth]{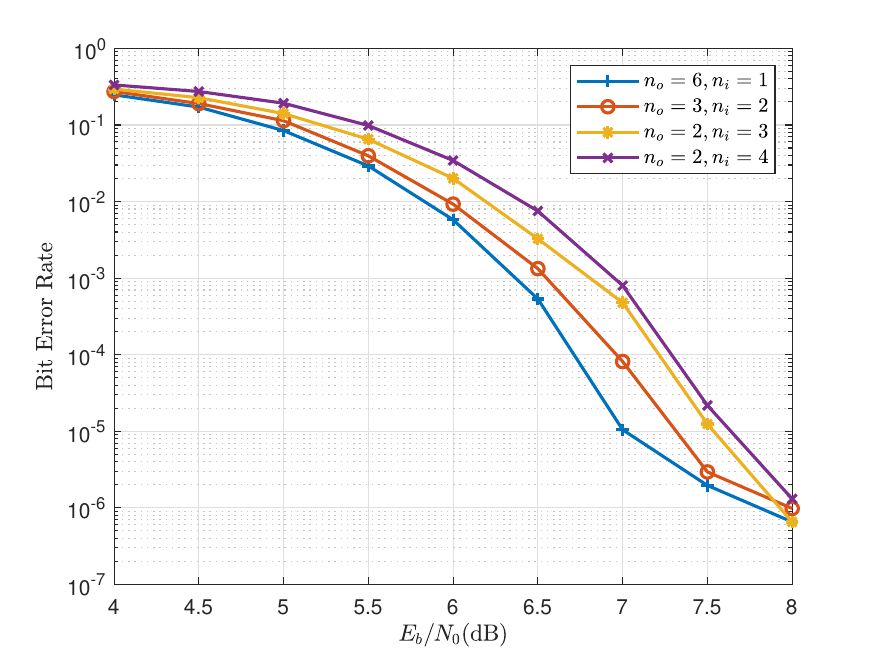}
    \caption{BER performance of the receiver with various numbers of inner iterations under the Proakis’ C channel.}
    \label{fig:6}
\end{figure}

Fig. \ref{fig:5} investigates the impact of varying the number of inner iterations $n_i$ on the algorithm's performance under the TU-6 channel, aiming to identify the optimal trade-off between computational complexity and accuracy. Based on the algorithm~\ref{algorithmic1}, it is necessary to clarify that the number of iterations in which the equalizer participates is $n_o$. On the other hand, the number of iterations involving the demodulator and decoder is $n_o \times n_i$. From Fig.~\ref{fig:5}, we find that when the number of inner iterations is set to $1$, $2$, $3$, and $4$, the number of outer iterations required for the algorithm to converge is $6$, $3$, $2$, and $2$, respectively. At the same time, the sub-figures show almost the same BER performance. Indeed, at an SNR of $4.5$dB, the BER for Fig.~\ref{fig:5}(a) to (d) is approximately $10^{-5}$. However, for Fig.~\ref{fig:5}(a) to (c), the number of iterations involved in both the demodulator and decoder is $6$. In contrast, the number of iterations in the equalizer is different, being $6$, $3$, and $2$ for Fig.~\ref{fig:5}(a), (b), and (c), respectively. This indicates that the cases in Fig.~\ref{fig:5}(b) and Fig.~\ref{fig:5}(c) require only one-half and one-third of the equalizer iterations, respectively, to achieve the same BER performance as Fig.~\ref{fig:5}(a).  In addition, when $n_i$ reaches $4$ in Fig.~\ref{fig:5}(d), the number of demodulator and decoder iterations increases to $8$, although the number of equalizer iterations remains $2$ as in Fig.~\ref{fig:5}(c). Consequently, increasing the number of $n_i$ can reduce the number of equalizer iterations. However, exceeding a certain threshold, the number of equalizer iterations no longer decreases, but the number of demodulation and decoding iterations increases. Therefore, for the TU-6 channel, the optimal choice of the inner iteration number is $n_i=3$, which achieves the same BER performance as shown in Fig.~\ref{fig:5}(a) with one-third of the equalizer iterations.\par 

In Fig. \ref{fig:6}, we vary the number of $n_i$ from $1$ to $4$ and investigate the BER performance of the algorithms in the Proakis’ C channel. The results show that in a degraded channel environment, the performance of the proposed algorithm gradually decreases as the value of $n_i$ increases. According to the analysis of Fig.~\ref{fig:5}, the proposed algorithm computational complexity is lowest when $n_i=3$, which corresponds to the yellow curve in Fig.~\ref{fig:6}. As can be seen from Fig.~\ref{fig:6}, $n_i=3$ has a performance loss of $0.5$dB compared to $n_i=1$ at a BER $10^{-5}$, but still has a gain of about $1$dB over the MMSE/SIC-based algorithm (see Fig.~\ref{fig:3}). Therefore, if computational resources are sufficient, we can choose $n_i=1$ for optimal performance, but if computational resources are limited, we can choose $n_i=2$ or $n_i=3$ to achieve a trade-off between performance and complexity.

\section{Conclusion}
In this paper, we investigate the problem of detecting CPM signals in the context of frequency-selective fading channels. We first present the CPM signal model and the corresponding factor graphs. Then, based on the BP rule and the UAMP, we propose an iterative equalization algorithm for CPM. The proposed algorithm converges more rapidly with fewer equalizer iterations with a designed the message schedule. Simulations demonstrate the effectiveness of the proposed algorithm, which outperforms the existing MMSE/SIC-based iterative algorithms by approximately $1.5$dB at a BER of $10^{-5}$ while maintaining a comparable complexity. 

\bibliographystyle{IEEEtran}
\bibliography{IEEEabrv,reference}

\end{document}